\documentstyle[12pt,a4]{article}

\begin{document}

\title{ Self-preservation of large-scale structures\\ in  Burgers'
turbulence\footnote{{\em To be submitted in Physics Letters A}} .}

\author{\large\bf  Erik Aurell $^{1,2}$, Sergey N. Gurbatov $^{3,4}$ \ \& Igor
I.  Wertgeim $^{2,5}$}
\maketitle

\bigskip

\begin{tabular}{ll}

$^1$ & Dept. of Mathematics, University of Stockholm\\
& P.O. Box 6701, S--113 85 Stockholm, SWEDEN\\\\
$^2$ & Center for Parallel Computers,\\
& Royal Institute of Technology,\\
& S--100 44 Stockholm, SWEDEN\\\\

$^3$ &Radiophysical Department, University of Nizhny Novgorod\\
&23, Gagarin Ave., Nizhny Novgorod, 603600, RUSSIA (Permanent address) \\ \\
$^4$ &Department of Mechanics,\\
& Royal Institute of Technology,\\
& S--100 44 Stockholm, SWEDEN\\\\

$^5$ & Institute of Continuous Media Mechanics,\\
& Russian Academy of Sciences,\\
& 1, Acad.~Korolyov Str., 614061, Perm, RUSSIA (Permanent address)
\end{tabular}

\date{}
\maketitle
\bigskip
\bigskip

	We investigate the stability of large-scale structures in Burgers' equation
under the  perturbation of high wave-number noise in the initial conditions.
Analytical estimates are obtained for random initial data with spatial spectral
density $k^n, n < 1$. Numerical investigations are performed for the case
$n=0$, using a parallel implementation of the Fast Legendre Transform.
\newpage

The appearance of ordered structures is possible for  nonlinear media with
dissipation\cite{Davis}. These structures often form a set of cells with
regular behaviour, alternating with randomly localised zones of dissipation.
One   nonlinear dissipative system  with such behaviour is  the well-known
Burgers' equation

\begin{equation}
	\frac{\partial v}{\partial t}+v\frac{\partial v}{\partial x}
=\mu\frac{\partial^{2} v}{\partial x^{2}} ;       \qquad     v(x,t=0)=v_{0}
(x)
\label{1}
\end{equation}
where $\mu$ is the  viscosity coefficient. The solutions of Burgers' equation
with random initial conditions display two mechanisms, which are also
 inherent to real turbulence:
  nonlinear transfer of energy through  the spectrum, and viscous damping in
the small scale region.
Equation (\ref{1}) was proposed by Burgers\cite{Bu2} as one-dimensional
model of fluid  turbulence. It has since been shown to arise
in a large variety of non-equilibrium phenomena,
when parity invariance holds\cite{Kura}.
Burgers' equation have
applications to nonlinear acoustics, nonlinear waves in thermoelastic media, to
modelling of the
formation of large-scale structures in the
Universe , and to many  other systems where  dispersion is negligibly small
compared  with nonlinearity (see \cite{GuMaSa} and references therein).

Burgers' turbulence is an example of strong turbulence,
the properties of which  are determined by the strong interaction of large
number of  harmonic waves.
	In Burgers' turbulence, due to the  coherent interaction of harmonics,
saw-tooth shock waves are formed,  which may be treated as a gas of strong
local interaction of  particles \cite{Tats,GuMaSa}. The collisions of shock
fronts leads to their merging, which is analogous to  inelastic collisions of
particles, and  to an increase of the integral  scale of turbulence.

	 In the limit of zero viscosity the solution of (\ref{1}) for the  velocity
field is given by

\begin{equation}
 v(x,t) =\frac{x-y(x,t)}{t}; \qquad  v(x,t)=-\frac{\partial S(x,t)}{\partial x}
\label{2}
\end{equation}
where $y(x,t)$ is where is realised the absolute maximum of the
function

\begin{equation}
 G(x,y,t)=S_{0}(y)-\frac{(x-y)^2}{2t}; \qquad  S_{0}(x)=\int^{x}v_{0}(y)dy
\label{3}
\end{equation}

By  analogy between the solutions of Burgers' equation and  the flow of
particles \cite{GuMaSa}, we shall call $S_{0}(y)$ the initial action, and
$S(x,t) = G(x,y(x,t),t)$ the  action at time $t$. This solution can be computed
by a
 Legendre transform of initial data \cite{SAF}.
We shall assume that the spectrum of random initial velocity field has a
following form:
\begin{equation}
g_{0}(k)=\alpha^{2}_{n} k^{n}b_{0}(k) \qquad b_{0}(k<k_{0})=1, \qquad
b_{0}(k>k_{0})=0
\label{4}
\end{equation}

The curvature $K_{s}$ of the initial action in (\ref{3}) may then be estimated
as
\begin {equation}
K^{2}_{s} =<(S_{0}^{\prime \prime}(y))^{2}>= \alpha^{2}_{n} k^{n+3}_{0}/(n+3) =
\sigma_{0}^{2}/l^{2}_{min}, \sigma_{0}^{2} =<v_{0}^{2}>=\alpha^{2}_{n}
k^{n+1}_{0}/(n+1)
\label{4.5}
\end{equation}

 Taking into account that the  curvature of parabola in (\ref{3})
is $1/t$, we get  that when $t\gg t_{n}=l_{min}/\sigma_{0}=K^{-1}_{s}$ the
curvature of the  parabola is much smaller than the curvature of the initial
action. This implies  that the global maximum of $G(x,y,t)$ is in the
neighbourhood of the  local maximum of $S_{0}(y)$, and that  $y(x,t)$ is a
stepwise  non-decreasing function of $x$. Thus the velocity field has a
universal  behaviour

\begin{equation}
 v(x,t) =\frac{x-y_{k}}{t} ; \qquad x_{k-1} < x < x_{k}
\label{5}
\end{equation}
in each cell between shock positions $x_{k}$. Each cell is characterized by two
numbers: the inverse Lagrangian coordinate  $y_{k}$, and the value of initial
action in the cell $S_{k} =S_{0}(y_{k})$. The dissipation zone is in this case
at the shock front, and hence have a zero width.
The positions of the shocks are determined by the equality at the point $x_{k}$
of the  values of two absolute maxima $G(x_{k},y_{k-1},t) =G(x_{k},y_{k},t)$:

\begin{equation}
 x_{k} =\frac{y_{k}+y_{k-1}}{2}+V_{k}t ; \qquad V_{k}=
\frac{S_{0}(y_{k-1})-S_{0}(y_{k})}{y_{k}-y_{k-1}}
\label{6}
\end{equation}
 It is easy to see that the rate of collisions of shocks  depends on the
asymptotic behaviour of the structure function of initial action
\begin{equation}
d_{s}(\rho)=<[S_{0}(x+\rho)-(S_{0}(x)]^{2}>\approx  \left\{
\begin {array}{ll}
 \alpha^{2}_{n}\rho^{1- n},  & \mbox{if  $n<1 $} \\
2\sigma_{s}^{2}, & \mbox{if $n>1 $}
\end{array}
 \right.
\label{7}
\end{equation}
Here $\sigma_{s}^{2} = <S_{0}^{2}>$.
 We can estimate the integral scale of turbulence $l(t)=|x-y|$ from the
condition that the parabola and the initial  action are of the same order:
\begin{equation}
d_{s}[l(t)]^{1/2}\approx l^{2}/t
\label{8}
\end{equation}
 From (\ref{7},\ref{8}) it therefore follows that  we have two different types
of growth for the scale $l(t)$:
\begin{equation}
l(t)\approx  \left\{  \begin {array}{ll}
 (\alpha t)^{2/(n+3)},  & \mbox{if  $n<1 $} \\
(\sigma_{s}t)^{1/2}, & \mbox{if $n>1 $}
\end{array}
 \right.
\label{9}
\end{equation}
It was shown \cite{GuMaSa}  (see also \cite{SAF}) that the common feature for
both types of initial spectra,
is the existence of self-similarity of statistical properties of solutions,
which are determined by only one scale $l(t)$.
{}From  (\ref{9}) one can see that in the case $n<1$, the behaviour of Burgers'
turbulence at  time $t$ is determined by the local large-scale behaviour of the
 initial spectrum. This  was the motivation for a  simple qualitative model of
Burgers' turbulence (at  $n<1$) as a discrete infinite  set of modes - the
spatial harmonics $k_{m}=k_{0}\gamma^{-m} (\gamma\gg 1)$,  sufficiently  spaced
in the spatial spectrum \cite{GuDeSa,GuCra}.
The amplitudes $A_{m}^{2}$ of harmonics were choosen from the condition that
the mean spectral density of harmonics  in the interval
$\Delta_{m}=k_{m+1}-k_{m}$ was identical to the random noise spectral density:
$A_{m}^{2}=g_{0}(k_{m})\Delta_{m}$. Assuming that the influence of
large-wavenumber components on large-scale one is small, it was  obtained that
the growth of
average scale for this discrete model is the same as in the case of continuous
spectrum given by (\ref{9}). The idea of independent evolution of large-scale
structures has also been used  to obtain  long time asymptotics of the
cylindrical   Burgers' equation  \cite{Enf}.

	It is possible to show that the assumption of relatively independent evolution
of the large-scale structures is valid for the continuous model of Burgers'
turbulence as well. In  \cite{GuDePr} it was obtained analytically that the
interaction of a regular positive  pulse with noise in  Burgers' equation does
not change the evolution of the  large-scale structure of the pulse, when the
value of initial action of the pulse is greater than the dispersion of the
noise action. This result was  confirmed by  numerical simulations based on the
asymptotic solution (\ref {2},\ref{3}) of Burgers' equation. Here we shall give
some simple estimations of this effect for the random perturbations with
initial spectrum (\ref{4}) and illustrate it by results of numerical experiment
for the case of truncated white noise (n=0 in (\ref{9})).

 Numerical experiments were performed using a parallel version of Fast Legendre
Transform alghorithm implemented  on a Connection Machine CM-200
\cite{SAF,AuWe}.
 The algorithm  uses  the property that the   inverse Lagrangian function
$y(x,t)$ is a non-decreasing function of  $x$, and  specific low-level
instructions on the Connection Machine, that allows one to perform
simultaneously    partial maximization operations over divisions of the range,
that are only known at run-time. The most primitive computation of the maximum
of (\ref{3}) takes  for $N$ points $O(N^{2})$ operations on a sequential
computer. The serial  Fast Legendre Transform takes  $O(N \log N )$ operations
in 1D (and  $O(N^2 \log^2 N)$ operations for data on a $N \times N$ grid in 2D
\cite{ Nul}). Our Parallel Fast Legendre Transform takes  $O(\log^2 N)$
operations on an ideal parallel computer with unlimited number of physical
processors, connected in hypercube network (this is the type of connection used
on the CM-2 and CM-200 models). The Connection Machine simulates an arbitrary
number of processors with a finite number (on our machine 8192). Our algorithm
will therefore eventually go linearly

 with the ratio of initialized gridpoints ($N$), to the actual number of
physical processors. For the class of one-dimensional problems we have
considered here, the computational time including input, output and Connection
Machine initialization is  much  less then one minute for total number of
points up to $N=2^{20}$.

	We shall consider the evolution of two initial random perturbations
$v_{0}(x)$ and $\widetilde{v_{0}}(x)$:
\begin{equation}
\widetilde{v_{0}}(x)=v_{0}(x)+  v_{h}(x)
\label{10}
\end{equation}
The power spectra of both processes  are described by (\ref{4}) with $n<1$, but
 the process $v_{0}(x)$  has spectral components in the range $ [0,k_{*}]$,
while the process  $\widetilde{v_{0}}(x)$ in the range  $ [0,k_{0}]$
with  $k_{0}\gg k_{*}$.
Restricted to the range  $ [0,k_{*}]$ the two processes are identical,
$v_{h}(x)$ being their
difference in the large wave-number range $[k_{*},k_{0}]$. The correlation
coefficient
between $v_0$ and $\widetilde{v_0}$ is $r_{0}=(k_{*}/k_{0})^{(n+1)/2}\ll1$. In
particular, we made calculations for $k_{0}^{(1)}/k_{*}=2^{2}$ and
$k_{0}^{(2)}/k_{*}=2^{6}$.

 	Three initial realisations, as described above, are shown on  Fig.1.
At $t\gg t_{n}\cong1/(\alpha k_{*}^{(n+3)/2})$ all three  processes are
transformed to a sequence of triangular pulses with the universal behaviour
inside the cells according to (\ref{5}). The solution $v(x,t)$  will be  stable
 relative to high wave-number perturbation $v_{h}(x)$, if the fluctuations of
both  the inverse Lagrangian coordinates  $\Delta  y_{k}=\tilde y_{k}-y_{k}$
and the  shock positions $\Delta x_{k}=\tilde x_{k}-x_{k}$, are small
respectively to integral scale of turbulence $l(t)$.
While the asymptotic properties of Burgers' turbulence are determined by the
behaviour of initial action, the values of disturbances $\Delta x_{k}$ and
$\Delta y_{k}$ will be determined by the value of dispersion  $\sigma_{hS}^{2}$
of perturbation action $S_{h}(x)= \int^{x}v_{h}(y)dy $:
\begin{equation}
\sigma_{hS}^{2}=<S_{h}^{2}>=\frac{\alpha_{n}^{2}}{(1-n)k_{*}^{1-n}}\left(1-\left(\frac{k_{*}}{k_{0}}\right)^{1-n}\right)                                         \label{11}
\end{equation}
and its correlation scale $l_{s}\approx(k_{*}k_{0})^{-1/2}$. The main
divergence between $\tilde v$ and $v$ is due to the different velocities of
shocks $V_{k}$ and $\tilde V_{k}$, and these errors increase in time. From
(\ref{6}) one  sees that the fluctuations of velocity are determined by the
fluctuations of the  action $S_{h}(y)$ in the neighbourhood of local maximum of
$S_{0}(y)$. This problem is similar to the analysis of statistical  properties
of Burgers' turbulence with stationary initial action \cite{GuMaSa}. Thus  we
get that the relative deviation of shock positions is equal to

\begin{equation}
\epsilon(t)=\frac{<(\Delta X_{k}^{2})>^{1/2}}{l(t)}\approx
\frac{\sigma_{hS}t}{l^{2}(t)}
                      \label{12}
\end{equation}
which  means that the behaviour of $\epsilon(t)$ depends on the growth law of
the integral scale $l(t)$. From (\ref{9},\ref{11},\ref{12}) we  get that in the
case $n<1$ and $k_{0}\gg k_{*}$:
\begin{equation}
\epsilon(t)\sim\frac{1}{(k_{*}l(t))^{\frac{1-n}{2}}}\frac{1}{\ln(\frac{k_{0}}{k_{*}})}\sim\frac{1}{t^{\frac{(1-n)}{(n+3)}}}                                     \label{13}
\end{equation}
Here we have taken  into account that for $k_{0}\gg k_{*}$, the value of
absolute maximum of the  random process $S_{h}(y)$ in the vicinity of local
maximum of $S_{0}(y)$, has a double logarithmic distribution with average value
$\sim \sigma_{hS}(\ln(k_{0}/k_{*}))^{1/2}$, and  dispersion decreasing with
correlation scale $l_{s}$,  proportional to $\ln(k_{0}/k_{*})$
(\cite{GuMaSa,Yak}). It may be that one shock in $v$ corresponds to a cluster
of shocks in $\tilde v$, but then they are spread out over a distance no more
than $\epsilon$ (\ref{13}).  Therefore one can say that the large scale
structures of Burgers' turbulence are self-preserving due to multiple merging
of shocks.
The stability of large-scale structures is illustrated by results of numerical
calculations, presented on Fig. 2. Three realisations of velocity field  are
shown, corresponding to ininitial conditions on Fig.1 with cutoff wavenumbers
equal respectively $k_{*}$,    $k_{0}^{(1)}$ and $k_{0}^{(2)}$.
One can see that the large-scale behaviour of all those realizations is
similar, and only fine structure weekly depends on cutoff wavenumber. The
correlation coefficient between the velocity fields with cutoff wavenumbers
$k_{*}$ and $k_0^{(2)}$ increase from the value $0.035$ for initial
perturbation (Fig. 1) up to the value $0.92$ on the stage of developed shocks
(Fig. 2). This become even more clear when comparing the   corresponding
initial and final actions, presented on Fig. 3 and Fig.4. There one  sees  that
the action for higher  cutoff values can be obtained from the smaller one
approximately by a shift along the  vertical axis. This is  in accordance
with estimations following from double logarithmic distribution of $S_{h}(y)$.

	In conclusion, we would like to stress that the stability of large-scale
structures relative to large wave-number perturbations,
is similar to the independence of asymptotic evolution of Burgers' turbulence
(with $n<1$)
of the value of viscosity coefficient $\mu $\cite{GuMaSa}.
We may thus interprete the action of small-scale perturbations on
solution of Burgers' equation as a turbulent viscosity, effectively changing
$\mu$, but leaving the initial conditions invariant \cite{GuSaYa,Yak}.

{\bf  \Large Acknowledgements}. This work is supported by the  G\"oran
Gustavsson Foundation
(S.N.G.), by the Swedish Institute (I.W.) and by the Swedish Natural Research
Council under contract S-FO 1778-302 (E.A.). The authors are grateful to the
Center for Parallel Computers
and the Department of Mechanics of the  Royal Institute of Technology for
hospitality.

\small{

}
\end{document}